\documentclass[prd,twocolumn]{revtex4}
\usepackage{dcolumn}
\usepackage{multirow}
\usepackage{graphicx}
\usepackage{amssymb}
\usepackage{bm}
\usepackage{hyperref}
\usepackage{epstopdf}
\usepackage{color}
\usepackage{mathrsfs}
\usepackage{amsmath,amssymb,amsthm}
\usepackage{rotating}
\usepackage{sverb, longtable}
\usepackage{subfigure}

\usepackage[graphicx]{realboxes}
\usepackage{adjustbox}

\begin{document}

\title{Updating neutrino mass constraints with Background measurements}

\author{Deng Wang}
\email{dengwang@ific.uv.es}
\author{Olga Mena}
\email{omena@ific.uv.es}
\affiliation{Instituto de F\'{i}sica Corpuscular (CSIC-Universitat de Val\`{e}ncia), E-46980 Paterna, Spain} 
\author{Eleonora Di Valentino}
\email{e.divalentino@sheffield.ac.uk}
\affiliation{School of Mathematics and Statistics, University of Sheffield, Hounsfield Road, Sheffield S3 7RH, United Kingdom}
\author{Stefano Gariazzo}
\email{stefano.gariazzo@unito.it}
\affiliation{Instituto de Fisica Teorica, CSIC-UAM
C/ Nicolás Cabrera 13-15, Campus de Cantoblanco UAM, 28049 Madrid, Spain}
\affiliation{Department of Physics, University of Turin, via P.\ Giuria 1, 10125 Turin (TO), Italy \looseness=-1}
\affiliation{Istituto Nazionale di Fisica Nucleare (INFN), Sezione di Torino, via P.\ Giuria 1, 10125 Turin (TO), Italy}

\begin{abstract}
Low-redshift probes, such as Baryon Acoustic Oscillations (BAO) and Supernovae Ia luminosity distances, have been shown to be crucial for improving the bounds on the total neutrino mass from cosmological observations, due to their ability to break degeneracies among the different parameters. Here, we expand background observations to include $H(z)$ measurements from cosmic chronometers, distance moduli from Gamma Ray Bursts (GRBs), and angular diameter distances from galaxy clusters. For the first time, using the physically motivated assumption of positive neutrino mass, we find that neutrino mass limits could be at 95\% CL below the minimal expectations from neutrino oscillation probes, suggesting possible non-standard neutrino and/or cosmological scenarios. Interestingly, it is not only the combination of the three background probes that is responsible for the $\sum m_\nu <0.06$~eV limits, but also each of them independently. 
The tightest bound we find here is $\sum m_\nu<0.043$~eV at 95\% CL after combining Cosmic Microwave Background Planck data with DESI BAO, Supernovae Ia, GRBs, cosmic chronometers, and galaxy clusters, showing a clear tension between neutrino oscillation results and cosmological analyses. In general, removing each one of three background probes still provides a limit $\sum m_\nu \lesssim 0.06$~eV, reassuring the enormous potential of these low-redshift observations in constraining the neutrino mass.
 
\end{abstract}
\maketitle

\section{Introduction}
\begin{figure*}
	\centering
	\includegraphics[scale=0.4]{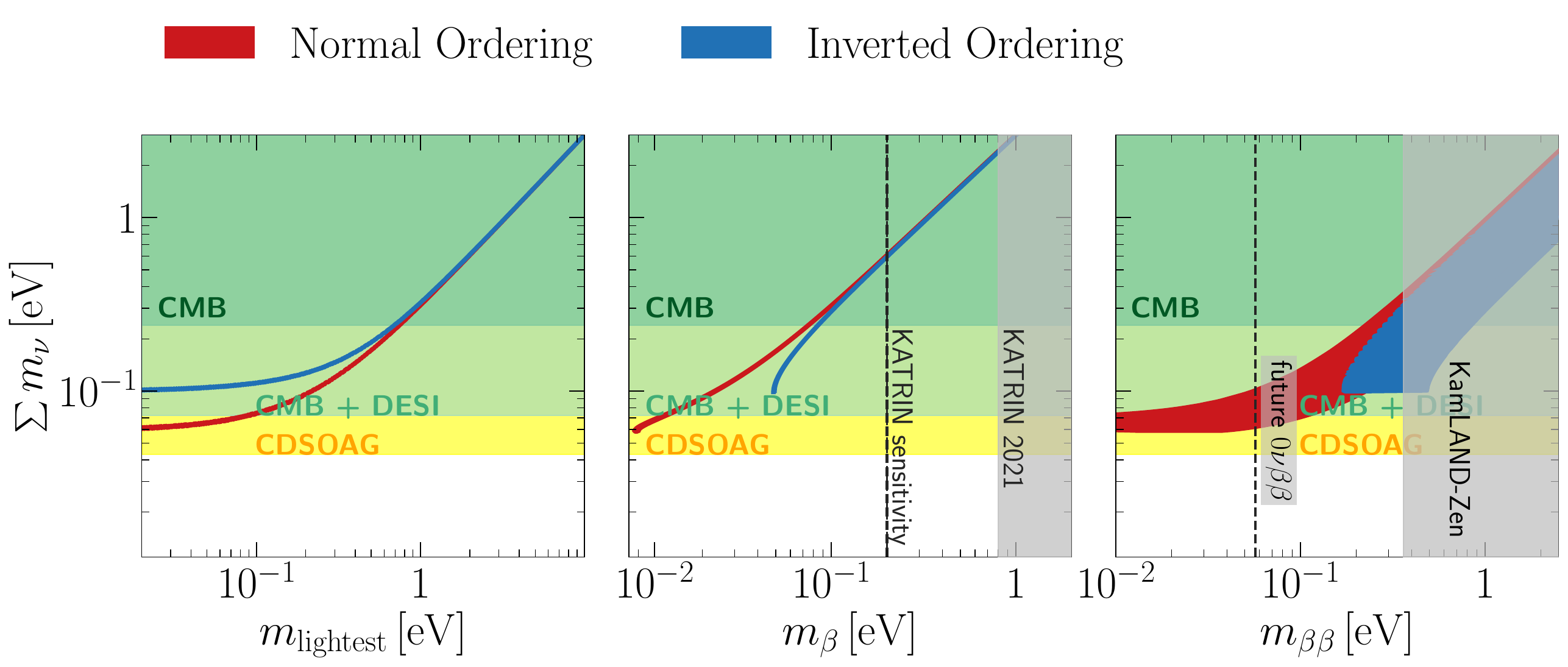}
    \caption{Sum of the neutrino masses ($\sum m_\nu$) as a function of other neutrino mass parameters: the lightest neutrino mass ($m_{\rm lightest}$, left panel), the effective beta-decay mass ($m_\beta$, middle panel), and the effective Majorana mass relevant for neutrinoless double beta decay probes ($m_{\beta\beta}$, right panel). Theoretical predictions for these quantities are depicted by the red and blue bands for normal (NO) and inverted ordering (IO), respectively. Cosmological constraints from CMB~\cite{Aghanim:2018eyx}, CMB+DESI~\cite{DESI:2024mwx}, and those derived here are depicted by the shaded green and yellow regions. Current and future $90\%$~CL sensitivities to $m_\beta$ can be found in Refs.~\cite{KATRIN:2021uub,Drexlin:2013lha}. Concerning neutrinoless double beta decay, current bounds constrain $m_{\beta\beta} < 0.036 - 0.156$ eV at $90\%$~CL~\cite{KamLAND-Zen:2022tow}, while next-generation experiments are expected to reach 3$\sigma$ sensitivities of $m_{\beta\beta} < 0.010 - 0.020$ eV~\cite{Agostini:2017jim,Giuliani:2019uno}.}\label{fig:fig1}
\end{figure*}

Sub-eV neutrinos are hot thermal relics. They could potentially constitute the entirety of the hot dark matter in our present universe, leaving clear signatures in cosmological observables (see, e.g.,~\cite{Lesgourgues:2018ncw,Lesgourgues:2006nd,Lattanzi:2017ubx,deSalas:2018bym,Vagnozzi:2019utt}), which can be exploited to place constraints on various neutrino properties, including their masses and abundances. Neutrino oscillation experiments have provided evidence for the existence of at least two massive neutrinos, as they measure two distinct squared mass differences, the atmospheric one $|\Delta m^2_{31}| \approx 2.55\cdot 10^{-3}$~eV$^2$ and the solar one $\Delta m^2_{21} \approx 7.5\cdot 10^{-5}$~eV$^2$~\cite{deSalas:2020pgw,Esteban:2020cvm}. Since the sign of $|\Delta m^2_{31}|$ remains unknown, two mass orderings are possible: \emph{normal} and \emph{inverted}. In the normal ordering, $\sum m_\nu \gtrsim 0.06$~eV, whereas in the inverted ordering, $\sum m_\nu \gtrsim 0.10 $~eV. It is intriguing that current cosmological constraints are surpassing the minimum sum of the neutrino masses allowed in the inverted hierarchical scenario (see Fig.~\ref{fig:fig1}). This suggests that cosmological observations could help in extracting the neutrino mass hierarchy~\cite{Gariazzo:2018pei,RoyChoudhury:2019hls,Hannestad:2016fog,Lattanzi:2020iik}, which is a crucial ingredient in future searches for neutrinoless double beta decay~\cite{Agostini:2017jim,Giuliani:2019uno}. 
Sub-eV neutrinos act as thermal relics with substantial velocity dispersions, leading to reduced clustering at scales smaller than their free streaming scale, consequently reducing the lensing power spectrum~\cite{Kaplinghat:2003bh,Lesgourgues:2005yv}. CMB bounds mostly rely on lensing: the Planck collaboration sets a bound $\sum m_\nu <0.24$~eV at $95\%$~CL from measurements of temperature, polarization, and lensing of the CMB~\cite{Aghanim:2018eyx} within the minimal $\Lambda$CDM model.

Nevertheless, it is precisely in the large-scale structure where the free streaming nature of neutrinos plays a significant role. For constraining the neutrino mass, the Baryon Acoustic Oscillation signature (BAO) is the typically exploited large-scale structure observable~\cite{Vagnozzi:2017ovm}. Along the line of sight direction, BAO data provide a redshift-dependent measurement of the Hubble parameter $H(z)$. Across the line of sight, BAO data can yield a measurement at the redshift of interest of the angular diameter distance, an integrated quantity of the expansion rate of the universe $H(z)$. Analyzing current cosmological data from the Planck CMB satellite, the SDSS-III and SDSS-IV galaxy clustering surveys~\cite{Dawson:2015wdb,Alam:2020sor}, and the Pantheon Supernova Ia sample, Ref.~\cite{DiValentino:2021hoh} established a very constraining neutrino mass bound, $\sum m_\nu<0.09$~eV at $95\%$~CL, see also Ref.~\cite{Palanque-Delabrouille:2019iyz}).
Current neutrino mass limits are challenging to circumvent within the $\Lambda$CDM framework and its simple extensions~\cite{diValentino:2022njd}, requiring the exploration of non-standard neutrino physics such as exotics beyond SM interactions or decays, and/or modified gravitational sectors.

Very recently the Dark Energy Spectroscopic Instrument (DESI) collaboration has presented new high-precision BAO measurements~\cite{DESI:2024uvr,DESI:2024lzq} and new cosmological results~\cite{DESI:2024mwx}. Combining DESI BAO measurements with CMB temperature, polarization, and lensing observations from the Planck satellite, as well as lensing observations from the Data Release 6 of the Atacama Cosmology Telescope (ACT)~\cite{ACT:2023kun,ACT:2023dou,ACT:2023ubw}, an upper bound $\sum m_\nu<0.072$~eV at $95\%$~CL is found~\cite{DESI:2024mwx}. 
Interestingly, an effective neutrino mass $\sum \tilde{m}_\nu \simeq -0.16 \pm 0.09$ eV is given in~\cite{Craig:2024tky} by extracting approximately the effect of massive neutrinos on the CMB lensing potential power spectrum. This result implies that current cosmological data may prefer the phenomenological effects that might correspond, through extrapolation, to a negative total neutrino mass.

In this study, we leverage both the new DESI BAO observations and the previous SDSS BAO observations, integrating them with CMB measurements and other background probes, including cosmic chronometers, galaxy clusters, and gamma-ray bursts. The impact of these background measurements on existing limits is remarkably profound, as we will demonstrate shortly. In many of the data combinations, \emph{the limit on the neutrino mass obtained in the standard scenario with stable neutrinos is smaller than the minimum expected from oscillation experiments, indicating a tension between cosmological and oscillation neutrino mass limits}.

This study is organized as follows. In the next section, we display the datasets and analysis methods. In Section III, we show our numerical results. Concluding remarks are presented in the final section.

\section{Data and methodology}

To investigate the neutrino sector on cosmological scales, we utilize the following observational datasets:
\begin{itemize}
\item CMB. We include the Planck 2018 high-$\ell$ \texttt{plik} temperature (TT) likelihood covering multipoles $30\leqslant\ell\leqslant2508$, polarization (EE), and their cross-correlation (TE) data spanning $30\leqslant\ell\leqslant1996$. Additionally, we incorporate the low-$\ell$ TT \texttt{Commander} and \texttt{SimAll} EE likelihoods within the range $2\leqslant\ell\leqslant29$~\cite{Planck:2019nip}. Moreover, we conservatively include the Planck lensing likelihood derived from \texttt{SMICA} maps across $8\leqslant\ell \leqslant400$~\cite{Planck:2018lbu}. This dataset is denoted as \textbf{``C''}.

\item SDSS. For the SDSS galaxy survey, we incorporate BAO measurements extracted from the 6dFGS~\cite{Beutler:2011hx}, MGS~\cite{Ross:2014qpa}, BOSS DR12~\cite{BOSS:2016wmc}, and eBOSS DR16~\cite{eBOSS:2020yzd,eBOSS:2020mzp} samples. 
This dataset is denoted as \textbf{``B''} hereafter.

\item DESI. For the DESI galaxy survey, we employ 12 DESI BAO measurements specified in Ref.~\cite{DESI:2024mwx} from various galaxy samples spanning the redshift range $0.1  < z < 4.16$~\cite{DESI:2024uvr,DESI:2024lzq,DESI:2024mwx}.
We refer to this dataset as \textbf{``D''}.

\item SN. The luminosity distances of type Ia supernovae serve as potent distance probes for investigating the background evolution of the universe, especially the equation of state of dark energy. In this analysis, we incorporate SN data points from the Pantheon+ sample~\cite{Scolnic:2021amr}, comprising 1701 light curves of 1550 spectroscopically confirmed type Ia Supernovae sourced from eighteen different surveys. Henceforth, we denote this dataset as \textbf{``S''}.

\item Cosmic Chronometers. This dataset is derived from analyzing the age variations among the most massive and passively evolving galaxies,
which can directly constrain the expansion history of the universe. Notably, the differential dating of passively evolving galaxies relies solely on atomic physics and does not involve integrating distances over redshift. Consequently, this probe remains independent of any specific cosmological scenario. Specifically, we utilize 31 data points from this dataset to help constrain the sum of neutrino masses~\cite{Moresco:2016mzx} at the background level. We refer to this dataset as \textbf{``O''}.

\item ADD.  Motivated by images from the Chandra and XMM-Newton telescopes, which show the elliptical surface brightness of galaxy clusters, the authors of~\cite{DeFilippis:2005hx} use an isothermal elliptical $\beta$ model to characterize the galaxy clusters. They constrain the intrinsic shapes of clusters to derive angular diameter distance (ADD) data by combining X-ray observations with Sunyaev-Zel'dovich (SZ) observations. The ADD data used in this analysis are derived from 25 galaxy clusters within the redshift range $z \in [0.023, 0.784]$~\cite{DeFilippis:2005hx}, from two subsets: 18 clusters from~\cite{Reese:2002sh} and 7 clusters from~\cite{2001ApJ...555L..11M}, where a spherical $\beta$ model was assumed. This dataset is denoted as \textbf{``A''}.

\item GRB. GRBs observations, originating from some of the most powerful sources in the universe, serve as valuable supplements at high redshifts for type Ia supernova observations.
These bursts, observed in the gamma-ray band, are nearly immune to dust extinction, enabling observations up to $z\sim$ 9~\cite{Salvaterra:2009ey,Cucchiara:2011hd}, far beyond the redshift range of observed type Ia supernovae ($z < 2.5$). Therefore, GRBs not only aid in exploring early universe physics but also offer high-$z$ constraints on the equation of state of dark energy at the background level. In this analysis, we employ calibrated distance measurements from 162 GRBs spanning the range $0.03 < z < 9.3$~\cite{Demianski:2016zxi} to constrain the sum of neutrino masses. We refer to this dataset as \textbf{``G''}.
\end{itemize}

To compute the theoretical power spectrum, we utilize the publicly available Boltzmann solver \texttt{CAMB}~\cite{Lewis:1999bs}. For the Bayesian analysis, we employ the Monte Carlo Markov Chain (MCMC) method to infer the posterior distributions of model parameters, leveraging the package \texttt{CosmoMC}~\cite{Lewis:2002ah,Lewis:2013hha}. We assess the convergence of the MCMC chains using the Gelman-Rubin statistics quantity $R-1\lesssim 0.02$~\cite{Gelman:1992zz} and analyze them using \texttt{Getdist}~\cite{Lewis:2019x}.

We specify uniform priors for the model parameters as follows: the baryon fraction $\Omega_bh^2 \in [0.005, 0.1]$, the cold dark matter fraction $\Omega_ch^2 \in [0.001, 0.99]$, the acoustic angular scale at the recombination epoch $100\theta_{MC} \in [0.5, 10]$, the scalar spectral index $n_s \in [0.8, 1.2]$, the amplitude of the primordial power spectrum $\ln(10^{10}A_s) \in [2, 4]$, the optical depth $\tau \in [0.01, 0.8]$, and the sum of masses of three active neutrinos $\Sigma m_\nu \in [0, 5]$ eV. We assume the degenerate neutrino hierarchy throughout this analysis.

\begin{figure*}
	\centering
	\includegraphics[scale=0.5]{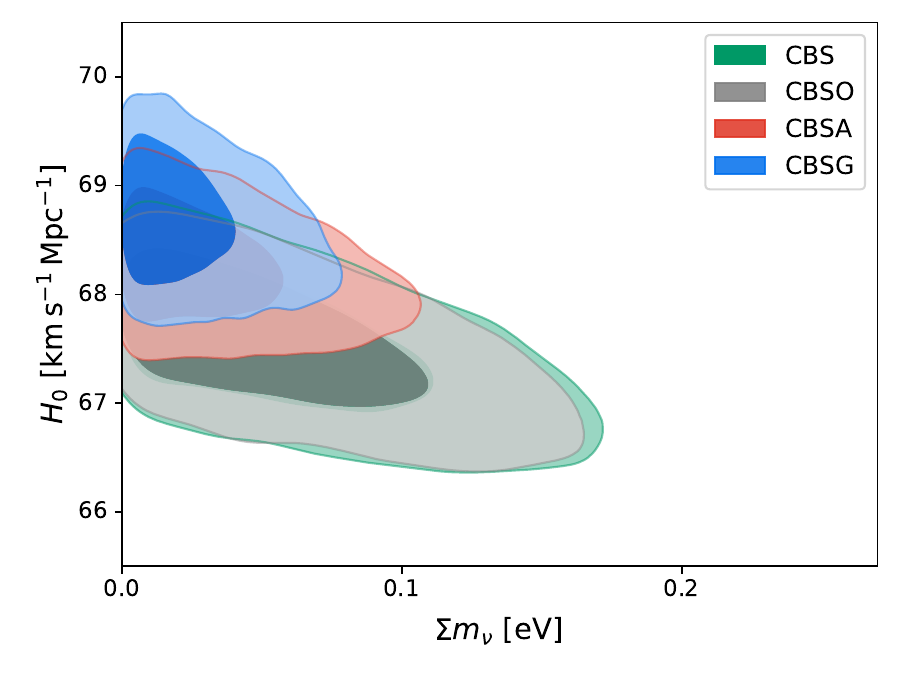}
        \includegraphics[scale=0.5]{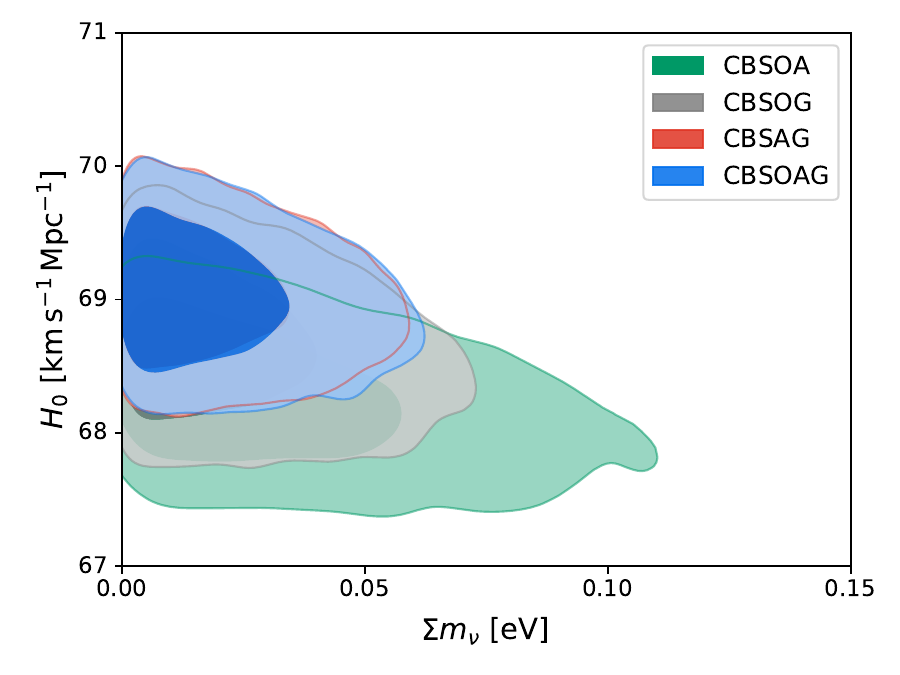}
	\caption{\textit{Left panel:} $68\%$ and $95\%$~CL limits in the ($\sum m_\nu$, $H_0$) plane for SDSS BAO combinations with CMB, SN, and the background measurements considered here, see the main text for details. \textit{Right panel:} as in the left panel but considering two or three background measurements in the combinations for the numerical analyses.}\label{fig:sdss}
\end{figure*}

\begin{figure*}
	\centering
	\includegraphics[scale=0.5]{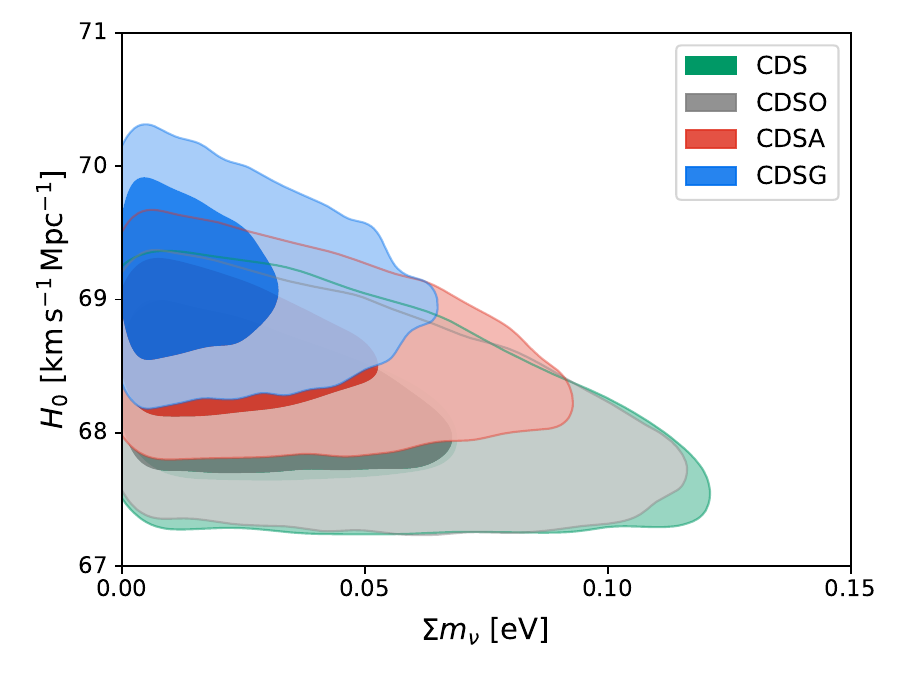}
        \includegraphics[scale=0.5]{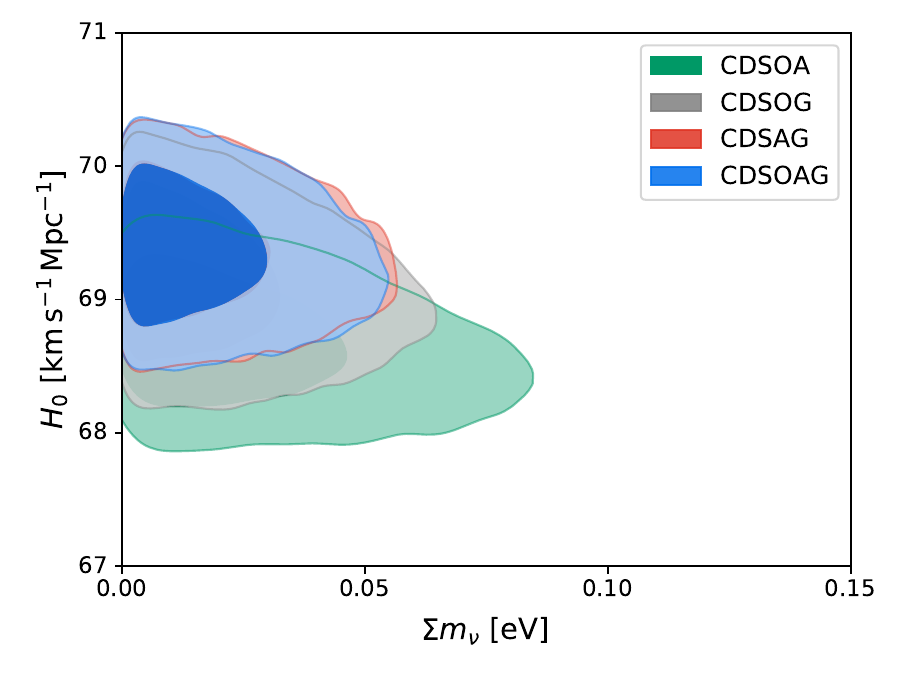}
	\caption{As Fig.~\ref{fig:sdss} but considering DESI BAO observations.}\label{fig:desi}
\end{figure*}

\begin{table*}[!t]
	\renewcommand\arraystretch{1.6}
	\setlength{\tabcolsep}{6mm}{
	\begin{tabular} { l |c| c |c| c }
		\hline
		\hline
		Datasets           &  $\Sigma m_\nu$ [eV]    &$H_0$ [km/s/Mpc]     &$\Omega_m$     &$\sigma_8$     \\
		\hline
		{\textbf{CBS}} & $<0.137$ \,\, ($2\,\sigma$)  &$67.54^{+0.52}_{-0.45}$    & $0.3130^{+0.0058}_{-0.0067}$  & $0.813^{+0.0110}_{-0.0075}$            \\
		
		{\textbf{CBSO}} & $<0.135$ \,\, ($2\,\sigma$)  &$67.56^{+0.51}_{-0.43}$    & $0.3128^{+0.0056}_{-0.0065}$  & $0.813^{+0.0110}_{-0.0077}$            \\
		
		{\textbf{CBSA}} & $<0.082$ \,\, ($2\,\sigma$)  &$68.27\pm 0.38$    & $0.3039\pm 0.0049$  & $0.8165^{+0.0081}_{-0.0066}$            \\
		
		{\textbf{CBSG}} & $<0.059$  \,\, ($2\,\sigma$) &$68.70\pm 0.43$    & $0.2991\pm 0.0054$  & $0.8184\pm 0.0068$            \\
  
            {\textbf{CBSOA}} & $<0.082$ \,\, ($2\,\sigma$)  &$68.28\pm 0.38$    & $0.3038\pm 0.0049$  & $0.8167^{+0.0081}_{-0.0063}$            \\
		
		{\textbf{CBSOG}} & $<0.056$ \,\, ($2\,\sigma$)  &$68.70\pm 0.42$    & $0.2991\pm 0.0053$  & $0.8185\pm 0.0068$            \\
  
            {\textbf{CBSAG}} & $<0.046$  \,\, ($2\,\sigma$) &$69.04\pm 0.37$    & $0.2949\pm 0.0047$  & $0.8177\pm 0.0065$            \\
		
		{\textbf{CBSOAG}} & $<0.049$ \,\, ($2\,\sigma$)  &$69.02\pm 0.37$    & $0.2951\pm 0.0047$  & $0.8179\pm 0.0068$            \\

            \hline
            
            {\textbf{CDS}} & $<0.093$ \,\, ($2\,\sigma$)  &$68.20\pm 0.44$    & $0.3045\pm 0.0056$  & $0.8156^{+0.0086}_{-0.0068}$            \\
		
		{\textbf{CDSO}} & $<0.091$ \,\, ($2\,\sigma$)  &$68.20\pm 0.43$    & $0.3045\pm 0.0054$  & $0.8158^{+0.0084}_{-0.0068}$            \\
		
		{\textbf{CDSA}} & $<0.071$ \,\, ($2\,\sigma$)  &$68.65\pm 0.37$    & $0.2990\pm 0.0047$  & $0.8160^{+0.0078}_{-0.0064}$            \\
		
		{\textbf{CDSG}} & $<0.049$  \,\, ($2\,\sigma$) &$69.17\pm 0.40$    & $0.2932\pm 0.0050$  & $0.8179\pm 0.0069$            \\
  
            {\textbf{CDSOA}} & $<0.065$ \,\, ($2\,\sigma$)  &$68.69\pm 0.36$    & $0.2984\pm 0.0044$  & $0.8166\pm 0.0071$            \\
		
		{\textbf{CDSOG}} & $<0.049$ \,\, ($2\,\sigma$)  &$69.14\pm 0.40$    & $0.2934\pm 0.0050$  & $0.8174\pm 0.0067$            \\
  
            {\textbf{CDSAG}} & $<0.045$  \,\, ($2\,\sigma$) &$69.38\pm 0.36$    & $0.2906\pm 0.0045$  & $0.8172\pm 0.0066$            \\
		
		{\textbf{CDSOAG}} & $<0.043$ \,\, ($2\,\sigma$)  &$69.38\pm 0.37$    & $0.2906\pm 0.0045$  & $0.8174\pm 0.0068$            \\

		\hline
		\hline
	\end{tabular}
	\caption{Mean values and uncertainties of four main cosmological parameters from different data combinations. Note that we quote 95\% CL errors for $\Sigma m_\nu$ and 68\% CL errors for $H_0$, $\Omega_m$, and $\sigma_8$. }
	\label{tab:bounds}}
\end{table*}

\section{Results}
\subsection{Constraining results}
Table~\ref{tab:bounds} summarizes the results obtained in this study. We start here with the constraints from SDSS BAO, Planck, and SN Ia observations. The combination \textbf{CBS} sets a 95\% CL bound on $\sum m_\nu<0.137$ eV. The addition of cosmic chronometers does not shift this constraint much. However, when angular diameter distances to galaxy clusters are considered (\textbf{CBSA}), the mean value of the Hubble parameter increases. Consequently, due to its anti-correlation with $\sum m_\nu$, the upper limit on this quantity is much tighter, $\sum m_\nu<0.082$ eV at 95\% CL. The addition of Cosmic Chronometers to this combination (\textbf{CBSOA}) does not change this limit.
Nevertheless, the most constraining bounds are obtained when GRB observations are added to the numerical analyses. In this case (\textbf{CBSG}), the value of the Hubble parameter is considerably larger, $H_0=68.70 \pm 0.43$ km/s/Mpc, and therefore the neutrino mass constraint is $\sum m_\nu<0.059$ eV at 95\% CL. Notice that this limit coincides with the minimum mass allowed by neutrino oscillation experiments. If we add data from galaxy clusters (\textbf{CBSAG}), the bound is $\sum m_\nu<0.046$ eV at 95\% CL due to the larger value of the Hubble constant obtained with this data combination.
Background measurements, by measuring the Hubble parameter, therefore have a strong potential to constrain the neutrino mass. If cosmic chronometers are added to the analysis (\textbf{CBSOAG}), we obtain $\sum m_\nu<0.049$ eV at 95\% CL. Needless to say, these limits are well below the standard expectations and would require physics beyond the standard model of particle physics, cosmology, or both.  Notice that the larger the mean value of the Hubble parameter, the smaller the bound on $\sum m_\nu$. As the neutrino mass constraint gets tighter, the amount of the matter mass-energy density gets smaller, as the product $\Omega_{\rm c} h^2$ cannot take very high values because of the CMB peaks location. 
To compensate for the low values of $\Omega_{\rm m}$, the clustering parameter $\sigma_{8}$ is increased. The most interesting case when considering SDSS BAO is the data combination \textbf{CBSOG}, as in this case, we can report a non-zero value for $\sum m_\nu$ at the $\sim$95\% CL, $\sum m_\nu =0.021^{+0.025}_{-0.021}$ eV.
It is also crucial to stress that
both galaxy clusters and GRBs are preferring a larger value of $H_0$, implying a lower bound on $\sum m_\nu$.
Figure~\ref{fig:sdss} shows the allowed two-dimensional contours in the ($\sum m_\nu$, $H_0$) plane for one, two, and three background probes added to the fiducial \textrm{CBS} data set.
Notice that 
all the values of the Hubble constant are consistent with each other at the $1-2\sigma$ level, ensuring the stability of our findings. This good level of agreement can be noticed from the results depicted in the supplementary material.

In the following, we shall discuss the bounds when DESI BAO observations replace the SDSS BAO dataset. Table~\ref{tab:bounds} shows that the limit without any of the background probes exploited here (\textbf{CDS}) is $\sum m_\nu < 0.093$ eV at 95\% CL, less constraining than that reported by the DESI collaboration due to the absence of the ACT lensing measurements in our numerical analyses. As in the SDSS BAO case, the data from cosmic chronometers do not change this limit significantly. Nevertheless, adding galaxy clusters (\textbf{CDSA}) shifts the limit down to $\sum m_\nu < 0.071$ eV at 95\% CL. 
GRBs prefer the neutrino total mass well below the (minimum) expectations from neutrino oscillation experiments.
The two-background data combinations \textbf{CDSOA} and \textbf{CDSAG} set 95\% CL limits of $\sum m_\nu < 0.065$ eV and $\sum m_\nu < 0.045$ eV respectively. As in the SDSS BAO case, the shifts in the limits of $\sum m_\nu$ are directly related to the smaller errors in the Hubble constant and in the matter mass-energy density as well as to the higher value of $H_0$ preferred by background measurements. Such high values of the Hubble constant, all consistent with each other at the $1-2\sigma$ level, are accompanied by low values of $\Omega_{\rm m}$ and large values of $\sigma_8$. Figure~\ref{fig:desi} shows the allowed two-dimensional contours in the ($\sum m_\nu$, $H_0$) plane for one, two, and three background probes added to the fiducial \textrm{CDS} data set, showing clearly the shift in $H_0$.
In Fig.~\ref{fig:fig1}, we illustrate via yellow regions the most constraining bound derived here, $\sum m_\nu < 0.043$ eV at 95\% CL, obtained in the case in which all three background probes are combined with CMB, DESI BAO, and SN Ia (\textbf{CDSOAG}). The three panels show the sum of the neutrino masses ($\sum m_\nu$) as a function of other neutrino mass parameters: the lightest neutrino mass ($m_{\rm lightest}$), the effective beta-decay mass ($m_\beta$), and the effective Majorana mass ($m_{\beta\beta}$),  relevant for neutrinoless double beta decay probes. We also depict the present $90\%$~CL sensitivities and forecasts for $\beta$-decay and neutrinoless double beta decay searches; see Refs.~\cite{KATRIN:2021uub,Drexlin:2013lha,KamLAND-Zen:2022tow,Agostini:2017jim,Giuliani:2019uno} for more details.

In the following, we present the constraints on the Hubble parameter and on the matter mass-energy density for each of the background observations exploited in this study, comparing them to those obtained by CMB observations to test their compatibility and justify the data combinations considered here.

Concerning cosmic chronometer data (\textbf{``O''}), the constraints are perfectly consistent with those obtained from CMB data alone ($H_0=67.36 \pm 0.54$~km/s/Mpc and $\Omega_{\rm m}=0.3153 \pm 0.0073$ for the CMB versus $H_0=67.8 \pm 3.1$~km/s/Mpc and $\Omega_{\rm m}=0.332^{+0.049}_{-0.069}$).

From galaxy clusters (\textbf{``A''}), we obtain $H_0=69.9 \pm 1.2$~km/s/Mpc and $\Omega_{\rm m}=0.317^{+0.075}_{-0.097}$. Even if the Hubble constant is slightly larger than that from Planck, it still remains consistent at the $\sim 2\sigma$ level. The value of the matter mass-energy density is instead in perfect agreement with that reported by Planck.

GBRs provide the largest value of $H_0$ ($77.7 \pm 7.5$~km/s/Mpc). However, the consistency with CMB observations is ensured due to the large error bars. The value of $\Omega_{\rm m}=0.52^{+0.11}_{-0.23}$ is also consistent with CMB values.

\begin{figure*}
	\centering
	\includegraphics[scale=0.6]{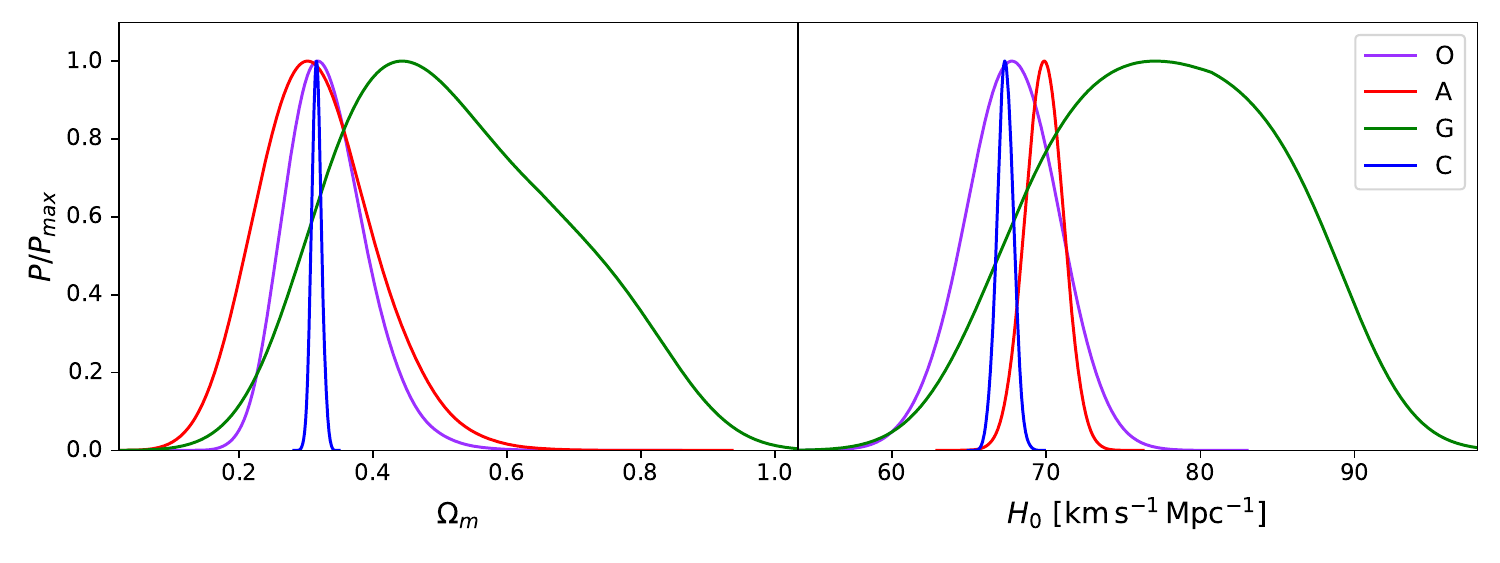}
	\caption{\textit{Left panel:} One-dimensional probability distribution for $\Omega_{\rm m}$ from each of the different datasets independently, where \textbf{``O''}, \textbf{``A''}, \textbf{``G''} and \textbf{``C''} refer to cosmic chronometers, galaxy clusters, GRBs and CMB respectively. 
 \textit{Right panel:} Same as in the left panel, but for $H_0$. }\label{fig:compare1}
\end{figure*}

\begin{figure*}
	\centering
	\includegraphics[scale=0.8]{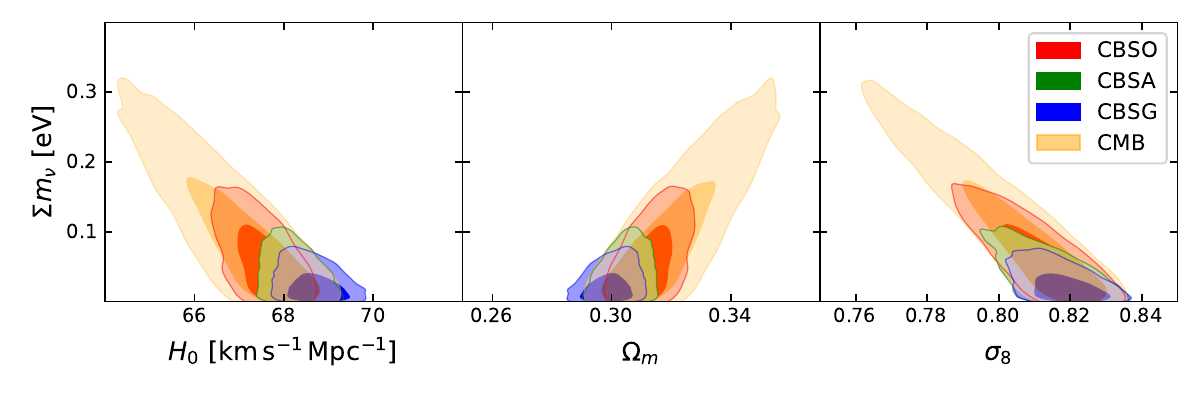}
        \includegraphics[scale=0.8]{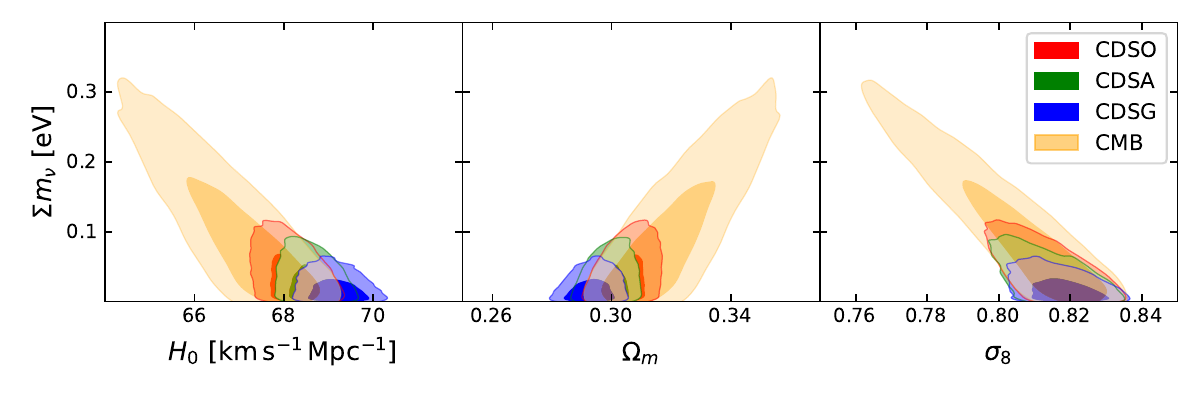}
	\caption{\textit{Upper panel}: Two-dimensional allowed contours in the ($H_0$, $\sum m_\nu$), ($\Omega_{\rm m}$, $\sum m_\nu$) and ($\sigma_8$, $\sum m_\nu$) planes arising from CMB alone (``CMB''), together with those obtained by adding each one of the background probes considered here independently plus SDSS BAO plus SN data. \textit{Lower panel}: Same as in the upper panel, but combining with DESI BAO and SN observations  (``CDS'').} \label{fig:compare2}
\end{figure*}

Figure~\ref{fig:compare1} depicts the one-dimensional probability distributions for $H_0$ and $\Omega_{\rm m}$ from each of these probes separately. Notice the overlapping of the different lines. This compatibility among the data sets justifies their combination. Figure~\ref{fig:compare2} shows the two-dimensional allowed regions in the ($H_0$, $\sum m_\nu$), ($\Omega_{\rm m}$, $\sum m_\nu$), and ($\sigma_8$, $\sum m_\nu$) planes for CMB alone and also after combining CMB observations with each of the background probes independently. As the value of the Hubble constant increases, the limit of $\sum m_\nu$ becomes tighter. The reduced errors on the matter mass-energy density also imply a lower bound on the neutrino mass. The overlapping in the allowed regions from each of the combinations justifies the constraints derived here and clearly demonstrates the robustness of our results.

\subsection{Tension between datasets}
When implementing a cosmological constraint, in general, there are two kinds of information that should be extracted. One is the constraining results from different data combinations, the other is the goodness of the fitting procedure. However, the consequence is different here. Since results are possibly affected by different datasets, we need to investigate the tension between datasets. We employ the effective number of standard deviations $n_\sigma \equiv \sqrt{2}\, \mathrm{Erf}^{-1}(P)$ \cite{Raveri:2021wfz} to characterize the tension between different datasets. Given an event of probability $P$, $n_\sigma$ corresponds to the number of standard deviations that an event with the same probability would have had were it drawn from a Gaussian distribution. This does not imply the Gaussianity of the underlying statistics and shall be understood as a logarithmic scale for probabilities. To estimate the discrepancies between datasets used in this study, we use the public package \texttt{tensiometer} \cite{Raveri:2021wfz}, which adopts a kernel density estimator (KDE) to compute the statistical quantity $n_\sigma$. Specifically, we take the bandwidth selector of the mean integrated square error to estimate the tension (see \cite{Raveri:2021wfz} for more details).

\begin{figure}
	\centering
	\includegraphics[scale=0.38]{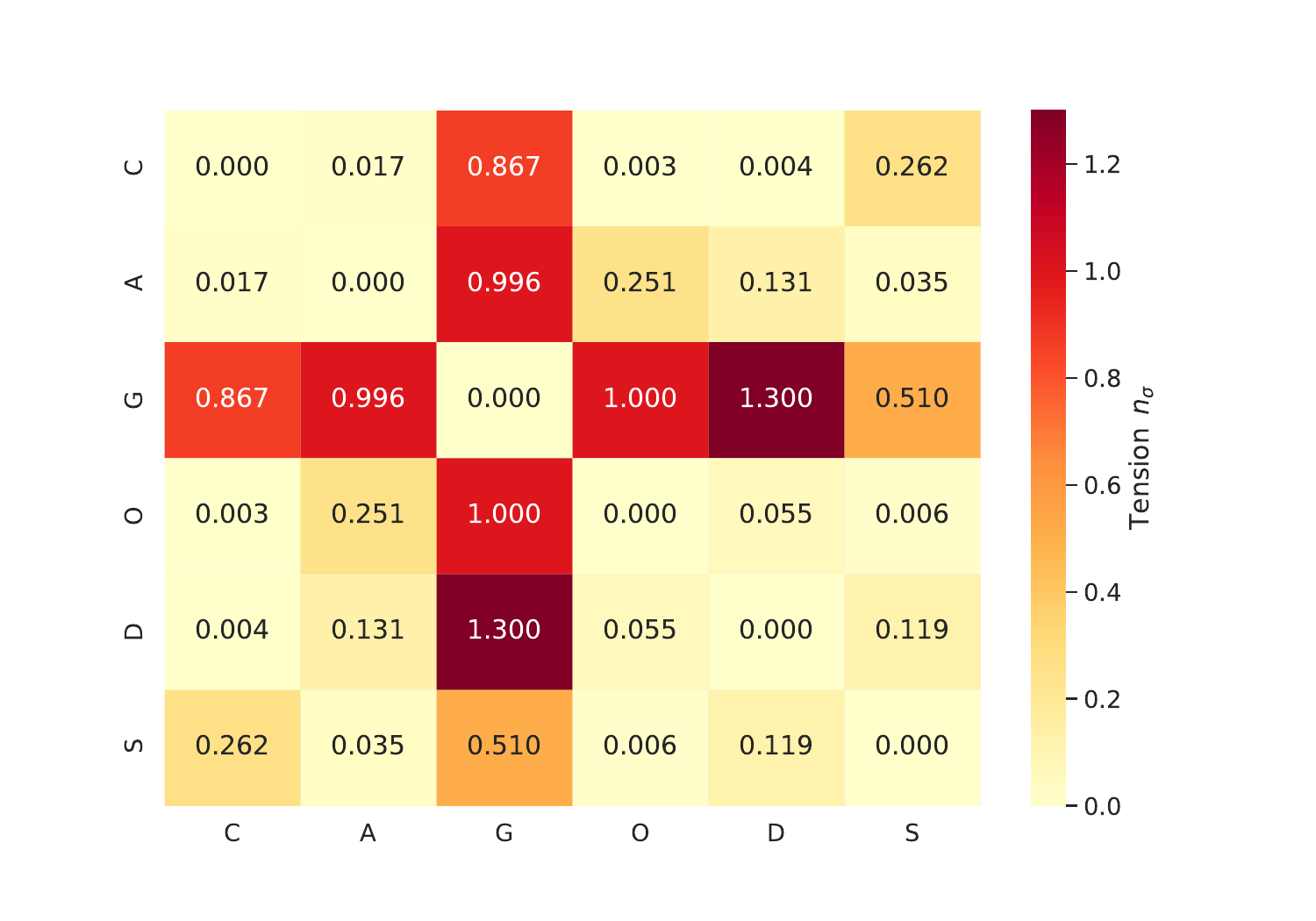}
	\caption{The tension (in number of standard deviations) between different datasets under the assumption of $\Lambda$CDM. Here C, A, G, O, D and S are CMB, ADD, GRBs, cosmic chronometers, DESI and SN, respectively. }\label{fig:heatmap}

 \end{figure}

We measure the tension between these six datasets in the framework of $\Lambda$CDM. The corresponding results are shown in Fig.~\ref{fig:heatmap}. We find that GRBs have an overall large tension with the other five datasets. Particularly, GRBs have the largest $1.3\,\sigma$ tension with DESI BAO data among the six datasets. The SN dataset has an overall small tension with other datasets. The smallest tension is $0.003\,\sigma$ between cosmic chronometers and CMB data. All the tension measurements are well within $1\,\sigma$, except for the $1.3\,\sigma$ inconsistency occurring between GRBs and DESI. This implies that our constraints on the neutrino mass sum are robust and would not be significantly affected by underlying tensions between the different datasets considered. Note that, while when we compute constraints on the cosmological
quantities we consider full CMB data including background and perturbation effects, for the purpose of evaluating the tensions between CMB and other datasets we only take into account the effect of CMB data at the background level.

\section{Conclusions}

Cosmological limits currently possess the highest constraining power on neutrino masses, if the standard scenario is assumed for neutrino decoupling and if neutrinos are stable particles with a constant mass. Albeit indirectly, these observations can also provide the first measurement of the neutrino mass ordering, by disfavoring inverted ordering more than normal ordering. 
In this manuscript, we use the new DESI BAO, 
and combine them with CMB data and other background probes, such as cosmic chronometers, galaxy clusters angular diameter distances, and gamma-ray bursts distance moduli. Very interestingly, several of the possible data combinations imply a neutrino mass limit smaller (at 95\% CL) than the minimum expected from oscillation experiments, i.e., $\sum m_\nu \lesssim 0.06$ eV, pointing towards a clear tension between cosmological and oscillation neutrino mass limits.
Therefore, despite the fact that such tension has not been detected so far~\cite{Gariazzo:2023joe}, not only one but several background probes indicate a clear problem between these two sets of observations. For instance, the combination of CMB with GRBs and SDSS BAO (DESI BAO) provides $\sum m_\nu <0.059$ eV ($\sum m_\nu < 0.049$) at 95\% CL.
If galaxy cluster observations are added to these measurements, the former limits are tightened to $\sum m_\nu <0.046$ eV and $\sum m_\nu <0.045$ eV respectively, both at 95\% CL. The most constraining bound we obtain is when CMB, SNIa, DESI BAO, and all background probes are combined:
this limit is $0.043$ eV at $95\%$~CL. Background probes are therefore able to provide strong bounds on the neutrino mass due to the preferred higher value of the Hubble constant and the smaller errors on both $H_0$ and the matter mass-energy density $\Omega_{\rm m}$.
Nevertheless, the values of $H_0$ obtained in the different dataset combinations exploited here are perfectly consistent within $1-2\sigma$. 
The bounds obtained here may imply the existence of very exotic cosmological scenarios (possibly related to the dark sector physics) and/or non-standard neutrino physics, since simple extensions of the $\Lambda$CDM model are not expected to significantly modify the bounds obtained here; see e.g.~\cite{diValentino:2022njd}.
Possible scenarios to relax the cosmological neutrino mass bounds include time-varying neutrino masses~\cite{Lorenz:2018fzb}, neutrino decay scenarios~\cite{Escudero:2020ped,Chacko:2020hmh,Chacko:2019nej}, long-range neutrino forces~\cite{Esteban:2021ozz}, neutrino cooling and heating~\cite{Craig:2024tky}, among others~\cite{Dvali:2016uhn}. Upcoming background measurements, such as those from future observations by DESI, will test all these possibilities while sharpening the cosmological neutrino mass limits.

\begin{acknowledgments}
DW is supported by the CDEIGENT Fellowship of Consejo Superior de Investigaciones Científicas (CSIC).
OM acknowledges the financial support from the MCIU with funding from the European Union NextGenerationEU (PRTR-C17.I01) and Generalitat Valenciana (ASFAE/2022/020).
EDV is supported by a Royal Society Dorothy Hodgkin Research Fellowship. 
SG is supported by the European Union’s Framework Programme for Research
and Innovation Horizon 2020 (2014–2020) under Junior Leader Fellowship LCF/BQ/PI23/11970034 by La Caixa Foundation
and by
the Research grant TAsP (Theoretical Astroparticle Physics) funded by Istituto Nazionale di
Fisica Nucleare (INFN).
This work has been supported by the Spanish MCIN/AEI/10.13039/501100011033 grants PID2020-113644GB-I00 and by the European ITN project HIDDeN (H2020-MSCA-ITN-2019/860881-HIDDeN) and SE project ASYMMETRY (HORIZON-MSCA-2021-SE-01/101086085-ASYMMETRY) and well as by the Generalitat Valenciana grants PROMETEO/2019/083 and CIPROM/2022/69. 
This article is based upon work from COST Action CA21136 Addressing observational tensions in cosmology with systematics and fundamental physics (CosmoVerse) supported by COST (European Cooperation in Science and Technology).
\end{acknowledgments}

\bibliographystyle{apsrev4-1}
\bibliography{main}

\end{document}